\documentclass[journal,draftcls,onecolumn,12pt,twoside]{IEEEtranTCOM}
\normalsize
\usepackage{amssymb,amsmath}
\usepackage{bbding}
\usepackage{flushend,cuted}
\usepackage{amsthm}
\usepackage{amsfonts}
\usepackage{float} 
\usepackage{srcltx}
\usepackage{mathrsfs} 
\usepackage{multirow}
\usepackage{amssymb}

\usepackage{graphicx}
\usepackage{epsfig}
\usepackage{psfrag}
\usepackage{subfigure}

\usepackage{setspace}
\usepackage{multicol}
\usepackage[noadjust]{cite}
\usepackage{hyperref}

\usepackage[]{units} 
\usepackage{url} 
\usepackage[dvips]{color}
\usepackage{verbatim} 
\usepackage{cite} 
\usepackage{ifthen} 
\usepackage{ifpdf} 
\usepackage{soul}
\usepackage{mathtools}

\newtheorem{theorem}{Theorem}

\newtheorem{remark}{Remark}

\newcommand*{\BODKIM}{}%

\makeatletter
\def\widebreve#1{\mathop{\vbox{\m@th\ialign{##\crcr\noalign{\kern3\p@}%
      \brevefill\crcr\noalign{\kern3\p@\nointerlineskip}%
      $\hfil\displaystyle{#1}\hfil$\crcr}}}\limits}

\def\brevefill{$\m@th \setbox\z@\hbox{$\braceld$}%
  \bracelu\leaders\vrule \@height\ht\z@ \@depth\z@\hfill\braceru$}
\renewcommand*\env@matrix[1][*\c@MaxMatrixCols c]{%
  \hskip -\arraycolsep
  \let\@ifnextchar\new@ifnextchar
  \array{#1}}
\makeatletter

\DeclareRobustCommand{\Nt}[1][Nt]{\ensuremath {N}}
\DeclareRobustCommand{\PrecMat}[1][Nt]{\ensuremath{G}}
\DeclareRobustCommand{\PowerMat}[1][Nt]{\ensuremath{P}}
\newcommand\figSize{0.5}

\DeclareMathAlphabet{\mathbsf}{OT1}{cmss}{bx}{n}
\DeclareMathAlphabet{\mathssf}{OT1}{cmss}{m}{sl}
\DeclareMathAlphabet{\mathcsf}{OT1}{cmss}{sbc}{n}

\newcommand{\svv}[1]{\mathbf{#1}}

\DeclareSymbolFont{bsfletters}{OT1}{cmss}{bx}{n}  
\DeclareSymbolFont{ssfletters}{OT1}{cmss}{m}{n}
\DeclareMathSymbol{\bsfGamma}{0}{bsfletters}{'000}
\DeclareMathSymbol{\ssfGamma}{0}{ssfletters}{'000}
\DeclareMathSymbol{\bsfDelta}{0}{bsfletters}{'001}
\DeclareMathSymbol{\ssfDelta}{0}{ssfletters}{'001}
\DeclareMathSymbol{\bsfTheta}{0}{bsfletters}{'002}
\DeclareMathSymbol{\ssfTheta}{0}{ssfletters}{'002}
\DeclareMathSymbol{\bsfLambda}{0}{bsfletters}{'003}
\DeclareMathSymbol{\ssfLambda}{0}{ssfletters}{'003}
\DeclareMathSymbol{\bsfXi}{0}{bsfletters}{'004}
\DeclareMathSymbol{\ssfXi}{0}{ssfletters}{'004}
\DeclareMathSymbol{\bsfPi}{0}{bsfletters}{'005}
\DeclareMathSymbol{\ssfPi}{0}{ssfletters}{'005}
\DeclareMathSymbol{\bsfSigma}{0}{bsfletters}{'006}
\DeclareMathSymbol{\ssfSigma}{0}{ssfletters}{'006}
\DeclareMathSymbol{\bsfUpsilon}{0}{bsfletters}{'007}
\DeclareMathSymbol{\ssfUpsilon}{0}{ssfletters}{'007}
\DeclareMathSymbol{\bsfPhi}{0}{bsfletters}{'010}
\DeclareMathSymbol{\ssfPhi}{0}{ssfletters}{'010}
\DeclareMathSymbol{\bsfPsi}{0}{bsfletters}{'011}
\DeclareMathSymbol{\ssfPsi}{0}{ssfletters}{'011}
\DeclareMathSymbol{\bsfOmega}{0}{bsfletters}{'012}
\DeclareMathSymbol{\ssfOmega}{0}{ssfletters}{'012}

\DeclareRobustCommand{\prob}[1][{\rm Pr}]{\ensuremath {{#1}}}

\begin{document}

\allowdisplaybreaks
%

\title{Diversity Combining via Universal Dimension-Reducing Space-Time Transformations}
\author{
\IEEEauthorblockN{Elad Domanovitz and Uri Erez}
\\
\IEEEauthorblockA{
Dept. EE-Systems, Tel Aviv University, Israel
}
}

\maketitle

\begin{abstract}
\ifdefined\BODKIM
Receiver diversity combining  methods play a key role in combating the detrimental effects of fading in wireless communication and other applications.
A novel diversity combining method is proposed where a universal, i.e., channel independent, orthogonal dimension-reducing space-time transformation is applied prior to quantization of the signals.
The scheme may be considered as the counterpart of Alamouti modulation, and more generally of orthogonal space-time block codes.
\fi

\end{abstract}

\section{Introduction}
In wireless communication, diversity methods play a central role in
combating the detrimental effects of severe channel variation (fading).
Of the many techniques that have been developed over the years with this goal, an important class involves the use of  multiple receive antennas. With sufficient separation between the antennas, each antenna may be viewed as a branch receiving the transmitted signal multiplied by an approximately independent fading coefficient. Diversity is achieved as the probability that the signal is severely affected by fading on all branches simultaneously is greatly reduced. The number of such (roughly) independent branches is commonly referred to as the diversity order.

Several methods of receive diversity combining are well known, most notably maximum-ratio combining (MRC), selection combining (SC), and equal-gain combining. All of these amount to performing a linear \emph{dimension-reducing} operation. For MRC, this operation is optimal in the sense of producing sufficient statistics, whereas in other methods some information loss is incurred in order to reduce some of the implementation prices inherent to MRC.

The dimension reduction aspect of the combining operation may serve several important goals. One important design goal is to reduce power consumption of a communication device. Most modern communication systems operate in the digital domain. Therefore, such a system must have at least one analog-to-digital conversion (ADC) unit that usually consumes a significant  amount of power. Some diversity combining methods, most notably SC, can serve to reduce the number of ADCs and thus result with power savings.

Another goal, that is also very relevant to modern communication systems, is reducing the bit rate of the digital interface between different digital blocks. For example, in a centralized (cloud) radio access network setting, each terminal (or relay) needs to be connected via a fronthaul link to the cloud. The combining  operation can be used to  reduce the required bit rate  when communicating over rate-constrained links.

In yet a different scenario, the dimension-reduction operation can serve to allow time-domain sub-Nyquist sampling.\footnote{We note that the connection between time-domain sampling and multiple-antenna receive combining is well known, e.g., the analogy between MRC and sampled matched filtering is clear.} When the  desired signal is known to posses some additional structure (beyond the frequency band it occupies), the sampling rate can be reduced significantly with a limited loss of information (due to noise accumulation, assuming the signal is contaminated by some noise); see, e.g.,  \cite{venkataramani2000perfect} and references therein. We will observe that certain scenarios of sub-Nyquist sampling can be recast as an equivalent multiple-antenna problem, and hence dimension-reducing diversity methods can play a role in
sub-Nyquist sampling.

As noted above, MRC is optimal in the sense of producing sufficient statistics. As recalled in more detail in the sequel, in MRC the outputs of different antennas are multiplied by channel-dependent weights. Since the combining depends on the specific channel realization, usually, this multiplication is carried out in the digital domain, in which case it requires sampling of all receive antennas, and employing an ADC for each.

Selection combining is an effective method to reap most of the benefits of MRC while reducing
the number of ADCs. Namely, in SC,
only the strongest receive antenna (or in hybrid schemes, a subset of antennas) is sampled.
This allows using less ADC modules, thus consuming less power, while paying only a small price in terms of performance.
A classical survey of receive diversity techniques is \cite{brennan1959linear}. More recent accounts
that also consider multiple-input multiple-output (MIMO) channels are \cite{molisch2004mimo,sanayei2004antenna}.

While SC is a practically appealing diversity-combining method in many applications, in common with  MRC, its implementation requires knowledge
(albeit, limited) of the channel in the selection phase. This requires implementing estimation and decision mechanisms that in certain scenarios may be prone to errors and add latency to the system. More importantly, as will be shown in the sequel, SC is ill-suited to scenarios
where multiple desired signals are
received, as will be most prominently
demonstrated for the case of a multiple-access channel.

We introduce a new linear diversity-combining scheme utilizing orthogonal space-time block codes. The key difference between the proposed scheme and
traditional linear combining schemes is that it is \emph{universal}. That is, the combining weights (in the proposed scheme, the space-time transformation) do not depend on the channel realization. As will be shown, in scenarios involving  multi-user detection, universal combining  has significant benefits over known linear combining schemes.

The rest of this paper is organized as follows. Section~\ref{sec:basic} describes the proposed method in the context of a wireless communication scenario with one transmit and two receive antennas. Section~\ref{sec:ADC} provides a performance comparison with known methods for both single-user and  multi-user scenarios. Section~\ref{sec:cloud} describes an application of the proposed method for relaying in a cloud radio access scenario.
Section~\ref{sec:dis} provides an extension of the method to more than two receive antennas. Section~\ref{sec:timeDomain} outlines the relation between the multiple-antenna scenario considered and sub-Nyquist sampling and demonstrates how the proposed method can be applied to the latter problem.

\ifdefined\BODKIM

\section{Description of the Scheme for Two Receive Antennas}
\label{sec:basic}
Consider a $2 \times 1$ single-input multiple-output (SIMO) channel, with channel coefficients $h_1$ and $h_2$, as depicted in
Figure~\ref{fig:basic secenario}.
\begin{figure}[htbp]
\begin{center}
\begin{psfrags}
\psfrag{a}[][][1]{$h_1$}
\psfrag{b}[][][1]{$h_2$}
\psfrag{x}[][][1]{$x$}
\includegraphics[width=\figSize\columnwidth]{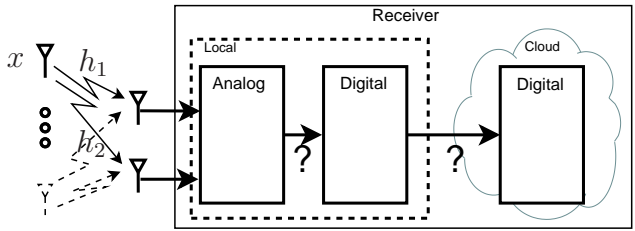}
\end{psfrags}
\end{center}
\caption{Basic scenario: receiver architecture for a $2 \times 1$ SIMO channel.}
\label{fig:basic secenario}
\end{figure}
The signal received at antenna $i=1,2$, at
discrete time $t$, is
\begin{align}
    s_{i}(t)=h_i x(t)+n_i(t).
    \label{eq:receive_sigs}
\end{align}
We assume that
the noise $n_i(t)$ is i.i.d. over space and time with
samples that are circularly-symmetric complex Gaussian random variables with unit variance. We further assume the transmitted symbols are subject to the power constraint $\mathbb{E}(|x|^2)=P$.

The scheme works on batches of two time instances and
for our purposes, it will suffice to describe it for
time instances $t=1,2$. Let us stack the four complex samples received over $T=2$ time instances, two over each antenna, into an $8 \times 1$
real vector:
\begin{align}
\mathbf{s}=[s_{1R}(1)  s_{1I}(1) s_{2R}(1) s_{2I}(1) s_{1R}(2)   s_{1I}(2)   s_{2R}(2)  s_{2I}(2)   ]^T,
\label{eq:s_vector}
\end{align}
where $x_R$ and $x_I$ denote the real and imaginary parts of a complex number $x$.
We similarly define the stacked noise vector $\svv{n}$.
Likewise, we define
\begin{align}
\mathbf{x}=[x_{R}(1) \, x_{I}(1)  \, x_{R}(2) \,  x_{I}(2)]^T.
\label{eq:source_vec}
\end{align}

Next, we form  a $4$-dimensional real
vector $\svv{y}$ by applying to the vector $\mathbf{s}$ the transformation
$\mathbf{y}=\svv{\PrecMat} \mathbf{s}$
where
\begin{align}
\svv{\PrecMat}=\frac{1}{\sqrt{2}}
  \left[ {\begin{array}{cccccccc}
   1 & 0 & 0 & 0 &  0 & 0 &  1 &  0 \\
   0 & 1 & 0 & 0 &  0 & 0 &  0 & -1 \\
   0 & 0 & 1 & 0 & -1 & 0 &  0 &  0 \\
   0 & 0 & 0 & 1 &  0 & 1 &  0 &  0
  \end{array} } \right].
  \label{eq:P}
\end{align}
Note that unlike conventional diversity-combining schemes, here the combining matrix
$\svv{\PrecMat}$ is \emph{universal}, i.e., it does not depend on the channel coefficients.

\begin{remark}
We note that the transpose of $\svv{\PrecMat}$ is precisely the description of the linear operation performed by Alamouti modulation \cite{alamouti1998simple} when expressed over the reals.
\end{remark}

It is readily shown that the following holds
\begin{align}
    \svv{y}&=\frac{\|\svv{h}\|}{\sqrt{2}}\svv{U}(h_1,h_2)\svv{x}+\svv{\PrecMat} \svv{n}  \nonumber \\
    &=\frac{\|\svv{h}\|}{\sqrt{2}}\svv{U}(h_1,h_2)\svv{x}+\svv{n'},
    \label{eq:n'}
\end{align}
where
\begin{align}
\svv{U}(h_1,h_2)=\frac{1}{\|\svv{h}\|}
  \left[ {\begin{array}{cccc}
  h_{1R} & -h_{1I}  &  h_{2R} & -h_{2I}  \\
  h_{1I} &  h_{1R}  & -h_{2I} & -h_{2R}  \\
  h_{2R} & -h_{2I}  & -h_{1R} &  h_{1I}  \\
  h_{2I} &  h_{2R}  &  h_{1I} &  h_{R1}
  \end{array} } \right].
  \label{eq:Ueff}
\end{align}
A key observation is that $\svv{U}(h_1,h_2)$ is an orthonormal matrix for any $h_1,h_2$:
\begin{align}
    \svv{U}^T(h_1,h_2)\svv{U}(h_1,h_2)= \svv{I},
    \label{eq:orthonormalityOfU}
\end{align}
where $\svv{I}$ is the identity matrix.
Further, since the rows of
$\svv{\PrecMat}$ are orthonormal, it follows that $\svv{n'}$ is i.i.d.
and Gaussian with variance $1/2$.\footnote{The variance is
$1/2$ as we chose above to normalize the complex noise to have unit power.}

We may reconstruct (up to additive noise)  the original samples by forming
\begin{align}
     \hat{\svv{x}} & = \svv{U}^T(h_1,h_2) \cdot \svv{y} \nonumber \\
     &=\frac{\|\svv{h}\|}{\sqrt{2}}\svv{x}+\svv{n''}
\end{align}
where $\svv{n''}$ is also i.i.d. Gaussian with variance $1/2$.


Since the dimension (over the reals) of $\svv{y}$ is four rather than eight, as is the dimension of the received signal $\svv{s}$, we obtained a universal dimension-reducing combining scheme.

We note that in order to perform the reconstruction, the channel
gains must of course be estimated as in true for any scheme, e.g, via the use of pilots. The difference is that in the proposed scheme, the estimation process occurs after the combining phase, i.e., using the effective MIMO channel \eqref{eq:n'}.

\section{Application to analog-to-digital conversion}
\label{sec:ADC}
In this section we demonstrate  the applicability of the scheme to analog-to-digital conversion  for power-limited receivers of narrowband signals. 
Similarly to SC, it  may be used to achieve maximal diversity order with a single radio-frequency (RF) chain and ADC.


We start by analyzing the performance in a scenario where a terminal with two antennas receives the signal transmitted from a single user equipped with a single antenna. In this scenario, setting aside hardware limitations, optimal SC outperforms the proposed method.
We then show, in contrast, that when the number of transmitting users increases, the new method is beneficial compared to all known methods that make use of a single RF chain.


\subsection{Single-User Scenario}



Consider again the scenario of a $2 \times 1$
SIMO system as depicted in Figure~\ref{fig:basic secenario} and described in the previous section. We note that as the fading coefficients are constants (rather than impulse responses), the model assumed is that of frequency-flat fading.


The best performance may be attained by quantizing (at sufficient resolution) the output of each antenna and then using MRC.
Applying MRC amounts to forming
\begin{align}
    y_{\rm MRC}&=\frac{1}{\|\svv{h}\|}\begin{bmatrix}h_1^* & h_2^* \end{bmatrix}\begin{bmatrix} {s}_1 \\ {s}_2\end{bmatrix}\nonumber \\
    &=\frac{\|h_1\|^2+\|h_2\|^2}{\|\svv{h}\|}x+\frac{h_1^*n_1+h_2^*n_2}{\|\svv{h}\|} \nonumber \\
    &=\underbrace{\|\svv{h}\|}_{\rm h_{eff,MRC}}x+n,
    \label{eq:eff_mrc}
\end{align}
where $n$ is white and Gaussian with unit variance.
This constitutes a sufficient statistic.
In particular, it is well known \cite{brennan1959linear} that  when $h_1$ and $h_2$ are independent, we obtain a diversity order of $2$.
The major downside of such a system is that two RF chains and ADCs are needed.

A classic alternative to MRC that requires only one RF chain and ADC is the method of  selection combining.
Here, rather than choosing the antenna arbitrarily, we choose the one with the higher signal-to-noise ratio (SNR).
Thus the effective channel becomes
\begin{equation}
y_{\rm SC}=\underbrace{\max(|h_1|,|h_2|)}_{\rm h_{eff,SC}} x +n,
\label{eq:eff_sc}
\end{equation}
where again $n$ is Gaussian noise with unit variance.
While the performance does not reach that of MRC, it  does attain
a diversity order of $2$.
The precise performance under independent  Rayleigh fading of SC is well known and may be found, e.g.,  in \cite{brennan1959linear}.

\begin{figure}[htbp]
\begin{center}
\begin{psfrags}
\psfrag{a}[][][1]{$h_1$}
\psfrag{b}[][][1]{$h_2$}
\psfrag{x}[][][1]{$\svv{x}$}
\psfrag{y1}[][][0.8]{$h_1x(2)$}
\psfrag{y2}[][][0.8]{$h_1x(1)$}
\psfrag{y3}[][][0.8]{$h_2x(2)$}
\psfrag{y4}[][][0.8]{$h_2x(1)$}
\psfrag{P}[][][1]{$\svv{\PrecMat}$}
\psfrag{U}[][][1]{$\svv{U}$}
\psfrag{z1}[][][0.8]{$\begin{bmatrix} y^1\\y^2\end{bmatrix}$}
\psfrag{z2}[][][0.8]{$\begin{bmatrix} y^3\\y^4\end{bmatrix}$}
\psfrag{j}[][][1]{$\hat{\svv{x}}$}
\psfrag{D}[][][0.7]{$\svv{D}$}
\includegraphics[width=\figSize\columnwidth]{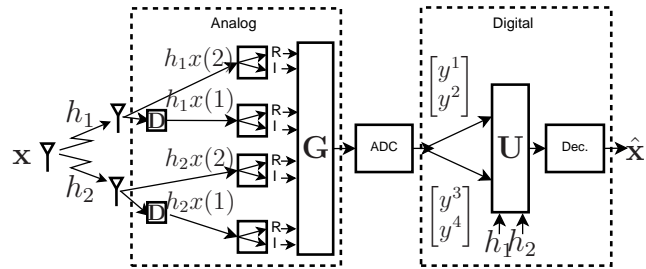}
\end{psfrags}
\end{center}
\caption{Proposed receiver front end employing a universal  orthogonal space-time diversity transformation.}
\label{fig:alamouti_quant}
\end{figure}

Alternatively, we may apply the space-time diversity combining method described in the previous section to the problem of ADC as follows. Since the processing matrix $\svv{\PrecMat}$ is fixed for all channels, it is possible to implement its operation in the analog domain (i.e., prior to quantization), requiring only delay, summation and negation elements.

As depicted in Figure~\ref{fig:alamouti_quant},
the received signals are first passed through the dimension-reducing transformation $\svv{\PrecMat}$ to obtain the vector $\svv{y}=
[   y^{1}    y^{2}    y^{3}    y^{4}]^T$
as defined in \eqref{eq:n'}
and
\eqref{eq:Ueff}.
Then, a (component-wise) scalar uniform quantizer $Q(\cdot)$ is applied to $\mathbf{y}$ to obtain $\svv{y}_q=Q(\mathbf{\mathbf{y}})$.
We denote the quantization error vector by
\begin{align}
    \svv{e} & = \svv{y}-\svv{y}_q \nonumber \\
    & =  \svv{y}-Q(\svv{y}).
    \label{eq:quanterror}
\end{align}

The sequence of quantized samples is  used to reconstruct an estimation of the source vector
\mbox{$\hat{\svv{x}} =
   [\hat{x}_R(1) \,
    \hat{x}_I(1)
    \,
    \hat{x}_R(2)
    \,
    \hat{x}_I(2)]^T
 $}
by applying the transformation:
\begin{align}
\hat{\svv{x}}
  &=\svv{U}(h_1,h_2)^T \svv{y}_q.
\end{align}
Using  \eqref{eq:n'} and \eqref{eq:quanterror}, we have
\begin{align}
\hat{\svv{x}}
  &=\svv{U}(h_1,h_2)^T(\svv{y}-\svv{e}) \\
  &=\svv{U}(h_1,h_2)^T\left(\frac{\|\svv{h}\|}{\sqrt{2}}\svv{U}(h_1,h_2)\svv{x} +\svv{n'} -\svv{e}\right) \\
  &=\underbrace{\frac{\|\svv{h}\|}{\sqrt{2}}}_{\rm h_{eff,Ala}} \svv{x} + \svv{n''} -\svv{e'},
\label{eq:eff_ala}
\end{align}
where $\svv{n}''$ has the same distribution as $\svv{n}$.

As for the quantization error $\svv{e}$ and its transformed variant $\svv{e'}$, we may invoke the standard assumption, that may be justified using subtractive dithered quantization, that it is independent of  the signal (and hence of $\svv{x}$) and is white (i.e., its covariance matrix is the scaled identity).

We conclude that the input/output relationship
of the proposed diversity combiner is identical
to that of  MRC, except for a power loss of a factor of two. In other words, we attain full diversity but no array gain, precisely as in the case of Alamouti
space-time diversity transmission. In comparison with SC (without taking into account implementation losses), there is a loss in the achieved SNR whereas an advantage is that no estimation of channel quality in the analog front end nor switching is required.


To compare the performance of the different schemes,
we first note that for all channel realizations
\begin{align}
|h_{\rm eff,Ala}|^2 \leq |h_{\rm eff,SC}|^2 \leq |h_{\rm eff,MRC}|^2
\end{align}
as these quantities correspond to the average, maximal, and sum of the (squared) channel gains, respectively.

We next compare the mutual information attained by each of the schemes in an  i.i.d Rayleigh fading environment. The mutual information is given by
\begin{align}
    I_{\rm scheme}(P)=\log\left(1+\PowerMat h^2_{\rm eff,scheme}\right),
\end{align}
where  $h_{\rm eff,scheme}$ is the effective scalar channel defined by either \eqref{eq:eff_mrc}, \eqref{eq:eff_sc} or \eqref{eq:eff_ala}.
As $h_{\rm eff,scheme}$ is a random variable, so is the resulting mutual information.

Figure~\ref{fig:alamouti_diversity_CDF_single} depicts the cumulative distribution function (CDF) of the mutual information attained by the three methods, for $P=1$ ($0$ dB). As can be seen, Alamouti combining outperforms arbitrary antenna selection for  ``bad'' channels but falls short of the performance of SC as expected.

Figure~\ref{fig:alamouti_diversity_Out_single} depicts the outage probability (assuming perfect coding) for a target rate of $2$ bits per complex symbol. As expected, Alamouti combining has a fixed $\sim3$ dB gap (factor of two power loss) from MRC, while the gap of SC is smaller.

\begin{figure}[htbp]
\begin{center}
\includegraphics[width=\figSize\columnwidth]{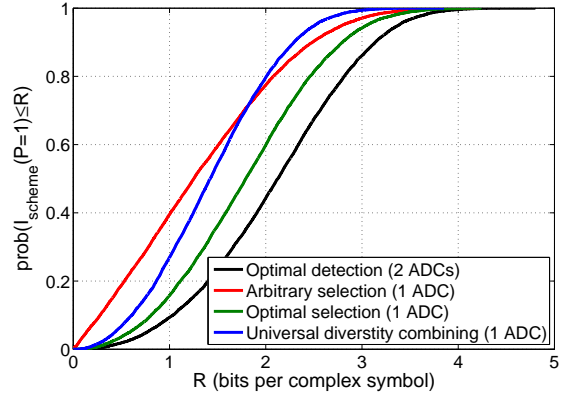}
\end{center}
\caption{Cumulative distribution function of the mutual information attained by different combining methods, for the case of single transmitter with a single antenna and a receiver with two receive antennas; i.i.d. Rayleigh fading is assumed with $P=1$.}
\label{fig:alamouti_diversity_CDF_single}
\end{figure}

\begin{figure}[htbp]
\begin{center}
\includegraphics[width=\figSize\columnwidth]{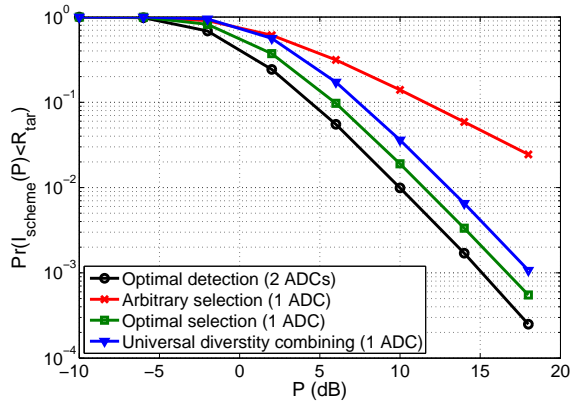}
\end{center}
\caption{Comparison of outage probability of different combining methods for an optimally encoded user transmitting over a $2 \times 1$ i.i.d. Rayleigh fading channel, with a target rate of $R_{\rm tar}=2$ bits per complex symbol.}
\label{fig:alamouti_diversity_Out_single}
\end{figure}

\subsection{Multi-User Scenario}
We consider now the scenario of a $2\times N$ MIMO-MAC system where $N$ users, each equipped with a single antenna, transmit to a common receiver that is equipped with two antennas. Again, it is assumed that  only a single RF chain is to be used at the receiver. As a figure of merit for performance, we now use the symmetric capacity (recalled below). We note that
unlike in the case of a single user,
sophisticated multi-user detection methods need to be applied in order to approach the considered figure of merit.


We first recall the more general channel model of a  MIMO-MAC with $N$ users, where each transmitter has $N_t$ antennas and the receiver has $N_r$ antennas. The input/output relation  can be expressed as
\begin{align}
    \svv{y}=\sum_{i=1}^{N}\svv{H}_i\svv{x}_i+\svv{n}
    \label{eq:MIMO_MAC}
\end{align}
where $\svv{H}_i$ is the channel matrix  between user $i$ and the receiver.
We assume isotropic (``white'') transmission by each user and that all users are subject to the same power constraint $P$.

Define a subset of users by $S \subseteq \{1, 2,\ldots, N\}$. Then,  the capacity region of the channel is given by (see, e.g., \cite{vishwanath2003duality}) all rate vectors $(R_1, \ldots, R_{N})$ satisfying
\begin{align}
\sum_{i\in S} R_i & \leq C(S) \nonumber \\
& \triangleq \log\det\left(\svv{I}+\PowerMat \sum_{i\in S}\svv{H}_i\svv{H}^H_i\right),
\label{eq:capacity_region_MAC}
\end{align}
for all subsets $S$ in the power set
of $\{1, 2,\ldots, N\}$.


If we impose the constraint that all users transmit at the same rate, then the maximal achievable rate is given by substituting $R_i=C_{\rm sym}/N$ in \eqref{eq:capacity_region_MAC}, from which it follows that the symmetric capacity is dictated by the bottleneck:
\begin{align}
C_{\rm sym}(\PowerMat)
&=\min_{S\subseteq\{1, 2,\ldots,\Nt\}}\frac{\Nt}{|S|}\log\det\left(\svv{I}+\PowerMat\sum_{i\in S}\svv{H}_i\svv{H}_i^H\right).
\label{eq:sym_capacity_MAC_ver2}
\end{align}

As all the combining methods considered involve only linear operations, we may obtain the associated symmetric capacity for each  by computing \eqref{eq:sym_capacity_MAC_ver2} for the respective effective channel.
We next derive explicitly the symmetric capacity associated with each method for the case of two users. 

We first consider the unrestricted symmetric capacity, i.e., the symmetric capacity for a system employing optimal reception (two RF chains).
We note that we may rewrite (\ref{eq:MIMO_MAC}) as
\begin{align}
    \svv{y}=\svv{h}_1\svv{x}_1+\svv{h}_2\svv{x}_2+\svv{n}
\end{align}
where
 \begin{align}
    \svv{h}_i=\begin{bmatrix}h_{1i} & h_{2i}\end{bmatrix}^T.
\label{eq:norm_hvec}
\end{align}
Denoting $\svv{H}_{\rm comb}=\left[\svv{h}_1~\svv{h}_2\right]$, \eqref{eq:sym_capacity_MAC_ver2} can be written as
\begin{align}
    C_{\rm sym,opt}(P)&=\min \left\{ C_{\rm opt}(\{1\}),C_{\rm opt}(\{2\}),C_{\rm opt}(\{1,2\})\right\} \nonumber \\
    & = \min\left\{2\log(1+\PowerMat\|\svv{h}_1\|^2),2\log(1+\PowerMat\|\svv{h}_2\|^2),\right. \nonumber \\ &~~~~~~~~ \left.\log\det\left(\svv{I}+\PowerMat \cdot \svv{H}_{\rm comb}\svv{H}_{\rm comb}^H \right) \right\}.
    \label{eq:csim_opt}
\end{align}
Similarly, for SC, the symmetric capacity can be expressed as
\begin{align}
    C_{\rm sym,SC}(P)&=\max_j\min \left\{ C_{\rm SC}(\{1\}),C_{\rm SC}(\{2\}),C_{\rm SC}(\{1,2\})\right\} \nonumber \\
    & = \max_j\min\left\{2\log\left(1+\PowerMat|h_{1,j}|^2\right),\right. \nonumber \\ &~~~~~~~~
    \left.2\log\left(1+\PowerMat |h_{2,j}|^2\right),\right. \nonumber \\ &~~~~~~~~ \left.\log\det\left(1+\PowerMat(|h_{1,j}|^2+|h_{2,j}|^2)\right) \right\}.
 \label{eq:csim_sc}
\end{align}

We now turn to the case  of the proposed method. By (\ref{eq:n'}) and \eqref{eq:MIMO_MAC}, the output is given by
\begin{align}
\svv{y}&=\sum_{i=1}^2\frac{\|\svv{h}_i\|}{\sqrt{2}}\svv{U}(h_{1i},h_{2i})\svv{x}_
i+
\svv{n}',
\end{align}
where $\svv{U}(h_{1i},h_{2i})$ is given by \eqref{eq:Ueff}.
Recalling that these matrices are orthonormal, we obtain
\begin{align}
    C_{\rm sym,Ala}(P)&=\min \left\{ C_{\rm Ala}(\{1\}),C_{\rm Ala}(\{2\}),C_{\rm Ala}(\{1,2\})\right\} \nonumber \\
    & = \min\left\{2\log\left(1+\frac{\PowerMat}{2}\|\svv{h}_1\|^2\right),\right. \nonumber \\ &~~~~~~~~
    \left.2\log\left(1+\frac{\PowerMat}{2}\|\svv{h}_2\|^2\right),\right. \nonumber \\ &~~~~~~~~ \left.\log\left(1+\frac{\PowerMat}{2}\left(\|\svv{h}_1\|^2+\|\svv{h}_2\|^2\right)\right)\right\}.
     \label{eq:csim_ala}
\end{align}

A comparison of the CDF of the symmetric capacity achieved by the different methods, in an  i.i.d. Rayleigh fading environment, is shown in Figures~\ref{fig:alamouti_diversity_CDF_multi} and \ref{fig:alamouti_diversity_Out_multi}.

Figure~\ref{fig:alamouti_diversity_CDF_multi} depicts the CDF of all three methods for $N=8$ users, each with power constraint $\PowerMat=1$. As can be seen, Alamouti combining clearly outperforms SC.

Figure~\ref{fig:alamouti_diversity_Out_multi} depicts the outage probability as a function of the SNR, where all users transmit at a common target rate of $2$ bits per complex channel use. While the universal combining scheme maintains the $\sim3$ dB gap from the MRC, SC suffers from a larger gap. In fact, it is easy to see that the (asymptotic in SNR) gap of SC becomes arbitrarily large as the number of users grows.

\begin{figure}[htbp]
\begin{center}
\includegraphics[width=\figSize\columnwidth]{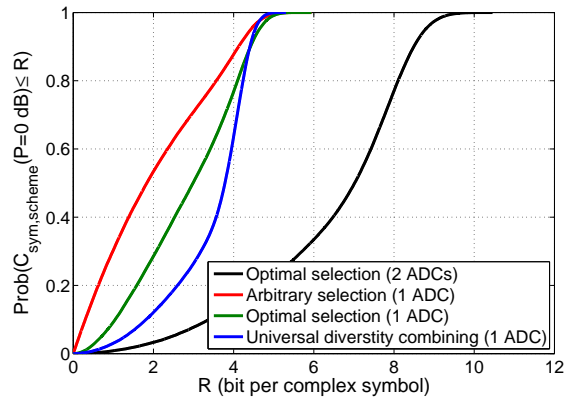}
\end{center}
\caption{CDF of the symmetric capacity associated with the different considered methods, in an i.i.d. Rayleigh fading channel environment with eight transmitters, each equipped with a single antenna and where the receiver is equipped with two antennas.}
\label{fig:alamouti_diversity_CDF_multi}
\end{figure}

\begin{figure}[htbp]
\begin{center}
\includegraphics[width=\figSize\columnwidth]{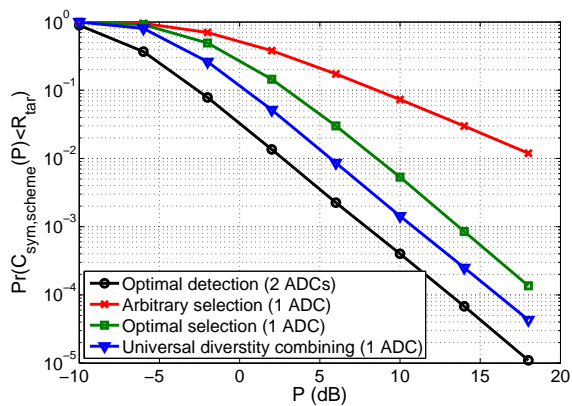}
\end{center}
\caption{Comparison of the outage probability associated with the different methods, in an  i.i.d. Rayleigh fading environment with eight transmitters, each equipped with a single antenna, and a common receiver equipped with two antennas. All users transmit at an equal rate $R_{\rm tar}$ such that $8R_{\rm tar}=2$ bits per complex symbol.}
\label{fig:alamouti_diversity_Out_multi}
\end{figure}

We may formalize the asymptotic performance of the universal diversity combining scheme in the form of a theorem.
\begin{theorem}
\label{thm:thm1}
For a Rayleigh fading $2 \times N$ MIMO-MAC, for any fixed (symmetric) target rate, at asymptotic high SNR, the universal combining scheme suffers a power penalty factor no greater than $2$ with respect to an optimal receiver.
\end{theorem}
We give the proof of the theorem for the case of two users in Appendix~\ref{sec:app1} where it is also explained how the general claim follows along similar lines.

\section{Application to ``Dumb'' Relaying for Multi-User Linear Detection at a Remote Destination}
\label{sec:cloud}
Another potential application of the proposed scheme is to be employed as part of a ``dumb'' relay. By a ``dumb'' relay we mean a relay (equipped with multiple antennas) that can only apply channel-independent linear processing to the antenna outputs followed by scalar quantization, the output of which is fed into a rate-constrained bit pipe.\footnote{This definition is similar to the definition of an instantaneous relay (see, e.g., \cite{el2007relay} and \cite{khormuji2010instantaneous}), with the additional requirement of linearity while allowing a small delay at the relay.}


Unlike in the previous section, the scheme we present now operates purely in the digital domain. A further difference is that we no longer assume frequency-flat fading. Rather, we will assume that after analog-to-digital conversion, a DFT operation is applied,  so that we are working in the frequency domain. In other words, the static channel we will consider is to be understood to apply to a single tone. The ``time" index $t$ will correspondingly refer to subsequent uses of the same tone, or in a practical setting could apply to adjacent tones as these typically have very similar channel coefficients.

We demonstrate the application to ``dumb" relaying in the context of the system described in
Figure~\ref{fig:two_double3}.
Here, two single-antenna users communicate with a central receiver via two relays, each equipped with two antennas, where the medium between the users and relays is a Rayleigh fading wireless channel, whereas the relays are connected to the central receiver via bit pipes.


The signal received at relay $i=1,2$ and antenna $j=1,2$ is given by
\begin{align}
    s^{i}_j(t)=h^i_{j1} \cdot  x_1(t) + h^{i}_{j2} \cdot x_2(t) +n^i_j(t),
    \label{eq:receive_sigs3}
\end{align}
and the corresponding channel matrix of relay $i$ is
\begin{align}
    \svv{H}^i= \left[ {\begin{array}{cc}
  h^i_{11} & h^i_{12}  \\
  h^i_{21} &  h^i_{22}
  \end{array} } \right].
\end{align}

\begin{figure}[htbp]
\begin{center}
\begin{psfrags}
\psfrag{h1}[][][1]{$\svv{H}^1$}
\psfrag{h2}[][][1]{$\svv{H}^2$}
\includegraphics[width=\figSize\columnwidth]{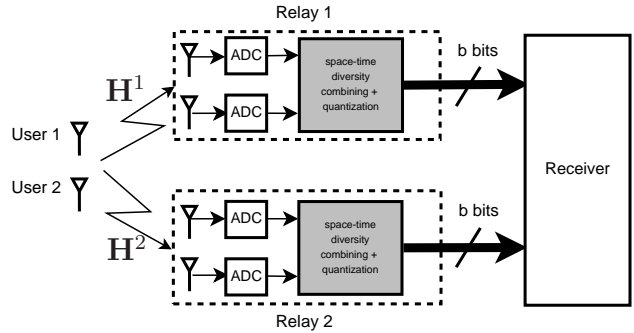}
\end{psfrags}
\end{center}
\caption{Two-user virtual MIMO system formed by two-antenna relays connected to a receiver via rate-constrained fronthaul links.}
\label{fig:two_double3}
\end{figure}

The question now arises
as to how best to utilize the finite number of bits available per sample in quantizing the output of the two antennas.
Note that not only do MRC and selection diversity depend on the use of channel state information (CSI) (which is precluded by the definition of a ``dumb" relay),
due to the distributed nature of the problem, both MRC and selection combining are also ineffective as the base station is interested in recovering both signals. 

Specifically, in order to perform useful CSI  dependent 
combining/selection at the relays, one would need to employ global channel state information, taking into account the channels from all users to all relays. 
For instance, to employ SC, the antenna selection at the relays would need to be performed jointly,  the methods proposed in 
\cite{heath2001antenna} (where linear equalization at the receiver was considered) being directly applicable. 
In the present context, such a process would need to take place in the cloud which would then notify each relay what combining/selection operation to employ. 
Such an approach, considering more general projection operations at the relays, has indeed been explored in the literature; we refer the reader to \cite{liu2015optimized} and references therein.



We now demonstrate that while keeping the bit rate fixed, one can benefit (albeit, not to the extent as with full CSI) from additional
antennas at the relays even without exploiting any channel state information at the relays.
Specifically, each relay can  provide diversity gains to \emph{both users} using the proposed diversity combining method, precisely since it makes no use of CSI at the linear combining stage, rather only in the reconstruction stage. 

Assuming both relays use the proposed space-time diversity combining scheme, the signal passed to the cloud from relay $i$ is given by
\begin{align*}
    \svv{y}^i&=\svv{U}(h^i_{11},h^i_{21})\svv{x}_1+ \svv{U}(h^i_{12},h^i_{22})\svv{x}_2+\svv{n'}^i,
\end{align*}
where $\svv{x}_j$ represents the real representation of the signal transmitted by user $j$
over the two time instances according to the notation
in \eqref{eq:source_vec}.
Thus, at the cloud we obtain the effective channel
\begin{align*}
  \left[ {\begin{array}{c}
  \svv{y}^1  \\
  \svv{y}^2
  \end{array} } \right] &=
  \underbrace{ \left[ {\begin{array}{c|c}
  \svv{U}(h^1_{11},h^1_{12}) &  \svv{U}(h^1_{21},h^1_{22}) \\
  \hline
  \svv{U}(h^2_{11},h^2_{12}) &  \svv{U}(h^2_{21},h^2_{22})
  \end{array} } \right]}_{\mathcal{G}} \left[ {\begin{array}{c}
  \svv{x}_1  \\
  \svv{x}_2
  \end{array} } \right]
  +
  \left[ {\begin{array}{c}
  \svv{n'}^1  \\
  \svv{n'}^2
  \end{array} } \right].
\end{align*}
Note that the effective matrix $\mathcal{G}$ has the desirable property that each of the four submatrices is orthogonal. Thus, it is expected that applying linear equalization to the effective channel followed by a slicer (or in general, a decoder) will exhibit some diversity gain.

The performance of the proposed scheme is demonstrated in Figure~\ref{fig:2Rel_2Ant_4_6_8_bits} for a simple scenario where the users transmit uncoded $16$-QAM symbols.\footnote{We chose to simulate uncoded
transmission to avoid the burden of computing the mutual information 
corresponding to quantized outputs. We believe that similar gains will
be manifested in coded transmission.}
We assume a simple receiver architecture that consists of linear equalization followed by single-user decoding. Employing such an architecture is reasonable since the in our example, the effective channel $\mathcal{G}$ is square (and well conditioned with high probability). As discussed in the previous section, when the number of transmitted streams is larger than the dimension of the received signal, multi-user detection techniques need be employed.
As a baseline for comparison, we consider a relay that quantizes and forwards the output of an arbitrary antenna; or alternatively, a relay that quantizes and forwards the output of both antennas but with half the number of bits allocated to each quantizer.\footnote{Since we assume ``dumb'' relays, the quantization of the inputs to the receiver was performed using a fixed (SNR independent) loading factor, taken as three times the standard deviation of the noise-free input to the quantizer.}
The latter is referred to as ``no combining" in Figure~\ref{fig:2Rel_2Ant_4_6_8_bits}.

Substantial improvement may be seen with respect
to the baseline schemes when a low bit error probability is desired, where we have considered quantization rates of $4$, $6$ and $8$ bits per sample for each relay.\footnote{As the gains are more pronounced at high SNR, we chose to demonstrate the performance of $16$-QAM rather than QPSK transmission.}


\begin{figure}[htbp]
\begin{center}
\includegraphics[width=\figSize\columnwidth]{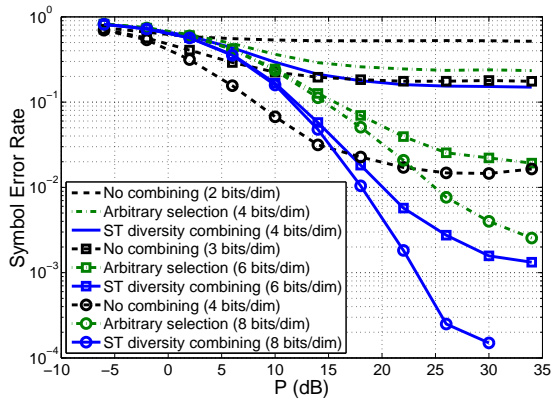}
\end{center}
\caption{Symbol error rate  achieved using ``dumb" relaying using the proposed diversity-combing scheme  and comparison to baseline relaying schemes for uncoded $16$-QAM transmission over a Rayleigh fading environment, where the receiver employs  MMSE equalization.}
\label{fig:2Rel_2Ant_4_6_8_bits}
\end{figure}

\section{Extensions to More than Two Antennas}
\label{sec:dis}
As in the case of space-time modulation for channel coding, extension of the scheme
to more receive antennas is possible, albeit with some loss.

A natural approach is to  try utilizing the theory of orthogonal designs.
It should be noted however that it is well known that the decoding delay (number of time instances stacked together) grows exponentially with the number of antennas. Another possible avenue is to try to follow the approach of quasi-orthogonal space-time codes
as developed in
\cite{tirkkonen2000minimal,jafarkhani2001quasi,sharma2003improved}. We demonstrate both approaches.

Attempting to apply orthogonal designs, one immediately confronts a basic obstacle due to the fact that rate-$1$ complex orthogonal designs do not exist beyond the case of two antennas \cite{tarokh1999space}. We next demonstrate the
problem that arises and also show how it may be resolved by judiciously combining balanced rate-$1/2$ orthogonal designs  \cite{adams2011novel} (which include the four basic OSTBCs described in \cite{tarokh1999space} for $2$-$8$ antennas) with repeated quantization used in conjunction with multiplicative dithering. For the sake
of concreteness and ease of exposition,
we demonstrate the method for the case of a SIMO system with $M=4$ receive antennas. 

The received signals are
given by \eqref{eq:receive_sigs}
where now \mbox{$i=1,\ldots,M$ (with $M=4$)}.
We proceed by stacking $T=8$ time instances of the received signal from the $4$ antennas and build an effective real-valued vector by decomposing each entry into its real and imaginary components, just as is done in \eqref{eq:s_vector}.
This yields for $M=4$, a vector $\svv{s}$ of dimension $2 \times 4 \times 8 =64$.

By reinterpreting the rate-$1/2$ orthogonal design of $4$ transmit antennas (see \cite{tarokh1999space}),
we arrive at a $8 \times 64$ transformation matrix $\svv{\PrecMat}$.\footnote{The specific form of $\svv{\PrecMat}$ can be found in Equation~(36) in \cite{domanovitz2017diversity}.}
Next, we form  a $8 \times 1$ real
vector $\svv{y}$ by applying to the effective received vector $\mathbf{s}$, formed in the manner described in \eqref{eq:s_vector}, the transformation
$\mathbf{y}=\svv{\PrecMat} \mathbf{s}$.

It can be shown that the following holds
\begin{align}
    \svv{y}&=\frac{\sqrt{2}\|\svv{h}\|}{\sqrt{8}}\svv{U}(h_1,h_2,h_3,h_4)\svv{x}+\svv{U} \svv{n}  \nonumber \\
    &=\frac{\|\svv{h}\|}{2}\svv{U}(h_1,h_2,h_3,h_4)\svv{x}+\svv{n'},
    \label{eq:n'_ext}
\end{align}
where $\svv{U}(h_1,h_2,h_3,h_4)$ is an $8 \times 16$ matrix with orthonormal rows.\footnote{The specific form of $\svv{U}(h_1,h_2,h_3,h_4)$ is given in Equation (29) in  \cite{domanovitz2017diversity}.} Here, the vector $\svv{x}$ is the $16$-dimensional real representation of the transmitted signal over $T=8$ time instances, formed analogously to
\eqref{eq:source_vec}.


Since the rows of  $\svv{U}(h_1,h_2,h_3,h_4)$ are  orthonormal, it follows that $\svv{n'}$ is white
(and Gaussian with  variance $1/2$).

The problem with using a non-rate $1$ orthogonal
design now becomes clear. Unlike $\svv{U}(h_1,h_2)$ (see \eqref{eq:Ueff}) which is square,
$\svv{U}(h_1,h_2,h_3,h_4)$ is non-square and hence is non-invertible.
We overcome this obstacle by passing the same
observation vector $\svv{s}$ via a ``dithered" version of $\svv{\PrecMat}$, such that another set of $8$ mutually orthogonal measurement rows is attained.
Specifically, let us define a $4$-dimensional vector
$\svv{d}=[d_1 \, d_2 \,  d_3 \, d_4]^T$ where $d_i$ are complex numbers of unit magnitude (pure phases). We form a dithered version
of the antenna outputs as
\begin{align}
    \tilde{s}_i(t)=d_i \cdot  s_i(t),
    \label{eq:tilde_s}
\end{align}
where $d_i$ does not depend on $t$. We assume that the $d_i$ are drawn at random as i.i.d. uniform phases.

We may associate with  $\tilde{s}_i(t)$, $t=1,\ldots,T=8$, the effective $64$-dimensional real vector $\tilde{\svv{s}}$.
Next, we  obtain another $8$-dimensional real
vector $\mathbf{\tilde{u}}$ by applying to the vector $\mathbf{\tilde{s}}$ the transformation $\mathbf{\tilde{y}}=\svv{\PrecMat} \mathbf{\tilde{s}}$.
We therefore obtain
\begin{align}
    \svv{\tilde{y}}&=\frac{\|\svv{h}\|}{2}\svv{U}(d_1 h_1,d_2 h_2,d_3 h_3,d_4 h_4)\svv{x}+\svv{n''},
    \label{eq:n''_ext}
\end{align}
where $\svv{n}''$ is distributed as $\svv{n'}$.

Note that the dithers  \eqref{eq:tilde_s} may be absorbed in $\svv{\PrecMat}$, thus defining a ``dithered" combining matrix $\PrecMat_{\rm dith}$.
Combining \eqref{eq:n'_ext} and \eqref{eq:n''_ext}, we
have
\begin{align}
  \underbrace{
  \left[ {\begin{array}{c}
  \svv{y}  \\
  \hline
  \svv{\tilde{y}}
  \end{array} } \right]}_{\svv{y_{\rm eff, dith}}} &=
  \underbrace{\frac{\|\svv{h}\|}{2} \left[ {\begin{array}{c}
  \svv{U}(h_1, h_2, h_3,h_4)  \\
  \hline
  \svv{U}(d_1 h_1,d_2 h_2,d_3 h_3,d_4 h_4)
  \end{array} } \right]}_{\mathcal{F}_{\rm dith}}
  \svv{x}
  +
  \left[ {\begin{array}{c}
  \svv{n'}  \\
  \svv{n''}
  \end{array} } \right].
  \label{eq:combined}
\end{align}
Finally, we apply component-wise quantization
to obtain
\begin{align}
\svv{y}_q=Q(\svv{y_{\rm eff, dith}}).
\end{align}
We may then recover an estimate of $\svv{x}$ by
applying the inverse of $\mathcal{F}$ to $\svv{y}$ or a linear MMSE estimator.

As mentioned above, another approach to extend the basic scheme to more antennas is to borrow ideas from quasi-orthogonal space-time codes. As an example for a quasi-orthogonal space-time linear combining matrix, we construct a matrix $\PrecMat_{\rm quasi}$ by taking half of the columns of
$\svv{\PrecMat}$, specifically columns
$1-16$ and $49-64$, scaling  by $\sqrt{2}$ to maintain orthonormality. This results in
\begin{align}
  \svv{y}_{\rm eff, quasi} &=
  \underbrace{\frac{\|\svv{h}\|}{2} \svv{U}_{\rm quasi}(h_1, h_2, h_3,h_4)}_{\mathcal{F}_{\rm quasi}}
  \svv{x}
  +
  \svv{n'}
  \label{eq:quasy}
\end{align}
where $\svv{U}_{\rm quasi}(h_1, h_2, h_3,h_4)$ is given by Equation~(35) in  \cite{domanovitz2017diversity}.

We tested the performance attained with both combining matrices in the scenario considered in
Section~\ref{sec:ADC}.
Specifically,
Figure~\ref{fig:alamouti_diversity4}
depicts the mutual information achieved  when using  different linear-combining schemes, for the case of a Rayleigh fading $4\times 1$ SIMO channel. 

We observe that both $\PrecMat_{\rm dith}$ and
$\PrecMat_{\rm quasi}$ achieve similar performance with some advantage to $\PrecMat_{\rm quasi}$.
Note that, in addition, $\PrecMat_{\rm quasi}$ utilizes only four consecutive symbols whereas $\PrecMat_{\rm dith}$  uses eight and hence induces less latency. On the other hand, the construction of $\PrecMat_{\rm dith}$ can be readily extended to more antennas.

We further observe that as both variants of space-time diversity combining do not achieve orthogonality, the gap is from MRC is larger than the minimal gap one could hope for (had orthogonality been possible) which is a factor of four in power loss ($\sim 6$ dB). 
Specifically, for $\PrecMat_{\rm quasi}$ the gap from MRC is roughly $7.5$ dB in the SNR range simulated. 
Similarly to the case of a two-antenna system,
the gap of SC to MRC is roughly half (in dB) that of the universal combininig scheme.

\begin{figure}[htbp]
\begin{center}
\includegraphics[width=\figSize\columnwidth]{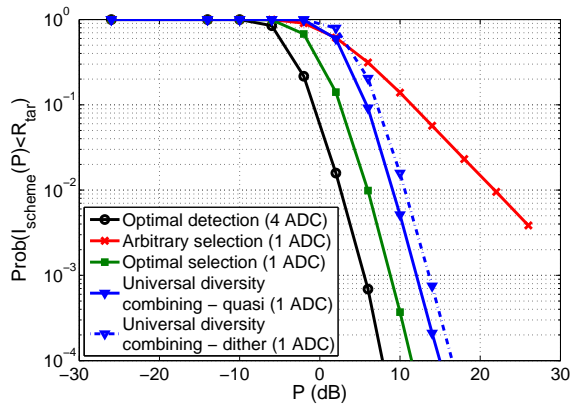}
\end{center}
\caption{Comparison of outage probability of different combining methods for an optimally encoded user transmitting over a $4 \times 1$ i.i.d. Rayleigh fading channel, with a target rate of $R_{\rm tar}=2$ bits per complex symbol.}
\label{fig:alamouti_diversity4}
\end{figure}

\section{Time-Domain Sub-Nyquist Interpretation/Application}
\label{sec:timeDomain}
In this section we build on the well-known analogy between
combining techniques for multiple-antenna arrays and those applied
for time-domain signals. For example, the relation between
maximal-ratio combining for antenna arrays and the sampled matched filter as an optimal front end (i.e., producing sufficient statistics) for a pulse-amplitude modulated time-domain signal is well recognized.

We describe how the developed diversity-combining technique may be leveraged to arrive at a sub-Nyquist signal acquisition method that is applicable to pulse-amplitude modulated signals.

Specifically, suppose we observe a (discrete-time) signal that is known to be of the from
\begin{align}
    \svv{s}(t)=\svv{h}x(t)+\svv{n}(t), \quad
    t=1, \ldots,K,
\end{align}
where $\svv{n}(t)$ is i.i.d. circularly-symmetric complex Gaussian noise of unit power and all vectors are column vectors of some dimension $N$. In other words, we know that the signal is sparse and lies in the $K$-dimensional subspace of $\mathcal{C}^{N K}$ spanned by
the vectors of the form $[\svv{0},\ldots,\svv{0},\svv{h}^T,\svv{0},\ldots,\svv{0}]^T$, as is the case in (the discrete-time representation of) pulse-amplitude modulation (PAM). We may view $\svv{h}$ as a pulse shape of length $N$.
The pulse shape used may change after $NK$ time instants.

For example, for the case of $K=3$, the basis assumes the form
\begin{align}
    \begin{bmatrix}
        \svv{h}     &  \svv{0} &  \svv{0}    \\
        \svv{0}     &  \svv{h} &  \svv{0}   \\
        \svv{0}     &  \svv{0} &  \svv{h}.
    \end{bmatrix}
\end{align}

An example for a scenario where the assumed model may be applicable to is  
sub-Nyquist detection of a frequency hopping signal. In a frequency hopping system, which is an effective method to combat jamming, the signal carrier is being chosen (based on a pseudo-random sequence) from a signal dictionary and is being changed at predefined symbol intervals.

As a concrete example, we may envision that $\svv{h}$ is a (complex) four-tap  carrier signal in a PAM transmission system that is chosen pseudo randomly from a ``dictionary''. Figure~\ref{fig:alamouti_diversity4} depicts  the real part of such a possible dictionary  consisting (in this example) of eight possible pulse shapes. The pulse shape chosen is kept constant for 
several symbols.



\begin{figure}[htbp]
\begin{center}
\includegraphics[width=\figSize\columnwidth]{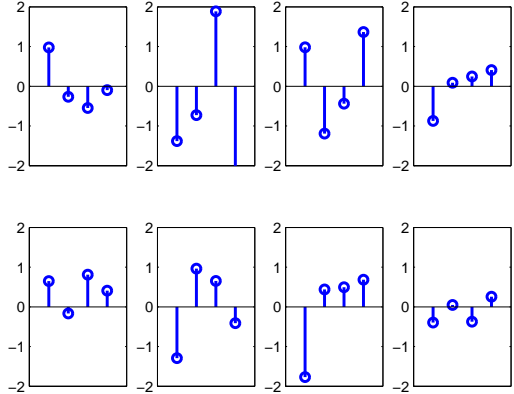}
\end{center}
\caption{Example of a pulse shape dictionary.}
\label{fig:alamouti_diversity4}
\end{figure}

For simplicity of exposition, we assume that the receiver is synchronized in the sense that it knows when each pulse shape starts and ends and also knows which member of the dictionary is being used.
Nonetheless, the proposed scheme may be advantageous in scenarios where  $\svv{h}$ is unknown at the time of signal acquisition (sampling) and is revealed to the receiver end (or is estimated by it) subsequently.


Figure~\ref{fig:option1MatchedFilter} depicts the optimal method for detection of the transmitted data. The signal is sampled at full rate.  Specifically, assuming the pulse shape occupies $T$ seconds in continuous time, sampling at full rate means sampling at a rate of  $\frac{N}{T}$ Hz (in the example, $N=4$). Then, a matched filter is applied and its output is sampled (at a rate of $1/T$ Hz) for data recovery.

We note that it is possible to implement the matched filter in the analog domain and sample at a rate of $1/T$ Hz. However, this requires implementing an analog filter bank, whose size should match that of the pulse shape dictionary and hence it is feasible only for small dictionaries.

\begin{figure*}[htbp]
\begin{center}
\begin{psfrags}
\psfrag{A}[][][1]{$\underbrace{}_\text{\rm T}$}
\includegraphics[width=1\textwidth]{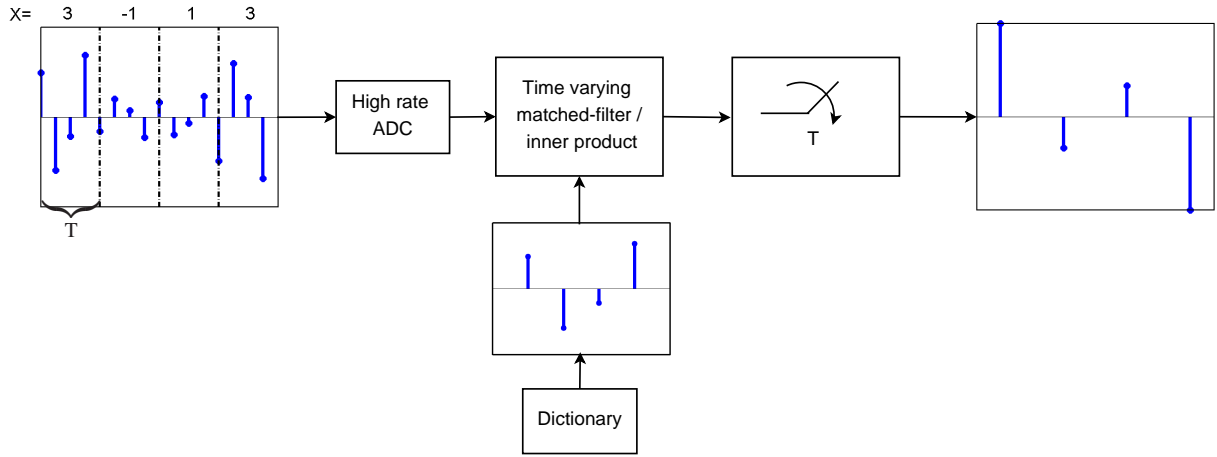}
\end{psfrags}
\end{center}
\caption{Optimal analog-to-digital conversion via a time-varying sampled matched filter.}
\label{fig:option1MatchedFilter}
\end{figure*}

The application of the proposed universal diversity method is demonstrated in Figure~\ref{fig:LowRateSampling}. In line with the duality mentioned above, the example considered where the pulse shape consists of four taps corresponds to a SIMO system with four receive antennas.


As was described in Section~\ref{sec:dis}, universal linear diversity combining is performed by applying an $8 \times 32$ precoding matrix $\PrecMat_{\rm quasi}$ (assuming quasi-orthogonal precoding is used). The precoding matrix is applied  to the stacked data of the real-valued representation of four consecutive PAM modulated complex symbols. We note that this requires implementing delay  as well as summation and negation elements in the analog domain. After sampling, the data is recovered, e.g., by applying linear MMSE equalization with respect to  $\mathcal{F}_{\rm quasi}$ followed by decoding.

We further note that one could also use tap selection (in analogy to antenna selection)  to reduce the sampling rate, i.e., sample at the time corresponding to the strongest tap of the pulse shape to achieve even better performance than that of universal diversity combining.
Nonetheless, whereas in the latter, the analog front end does not vary in time, optimal selection translates to applying a shift in the sampling time, every time the pulse shape changes.

Finally, we note that whereas when one considers an antenna array in a wireless fading environment, it is usually hard to expect that the channel coefficients remain constant over many symbols, thus precluding the use of space-time block codes when the number of antennas is large (as the needed coherence time grows exponentially with the number of antennas). In contrast, when considering the application to a time-domain signal, it is very reasonable to assume that the pulse shape remains constant over a long period of time and as a consequence, one can apply the proposed scheme to pulse shapes consisting of many taps (at the expense of considerable processing complexity in the digital domain).



\begin{figure*}[htbp]
\begin{center}
\begin{psfrags}
\psfrag{A}[][][1]{$\underbrace{}_\text{\rm T}$}
\includegraphics[width=1\textwidth]{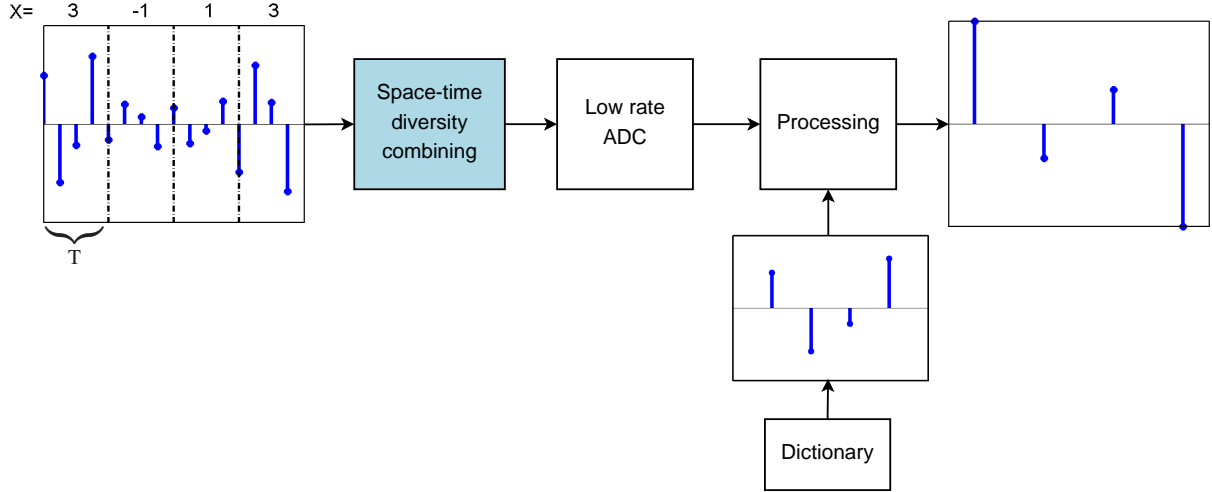}
\end{psfrags}
\end{center}
\caption{Sub-Nyquist sampling via space-time diversity combining.}
\label{fig:LowRateSampling}
\end{figure*}

\fi

\appendix
\section{Proof of Theorem~\ref{thm:thm1}}
\label{sec:app1}



We start by establishing the limit
\begin{align}
    \lim_{P \rightarrow \infty} \frac{\prob\left(C_{\rm sym,Ala}(P)<R\right)}{\prob\left(C_{\rm sym,opt}(P)<R\right)}=16,
    \label{eq:limAla}
\end{align}
where  fixed $R$ is any (fixed) target rate and where we consider the case of two users. We then observe that such an increase in outage probability corresponds to no more than a factor of two in terms of power penalty.

With some abuse of notation we now make the dependence of $C(S)$ on $P$  (as defined in \eqref{eq:capacity_region_MAC}) explicit and denote
it by $C(S)=C(S,P)$.

To show that \eqref{eq:limAla} holds, we show that
\begin{align}
    \limsup_{P \rightarrow \infty} \frac{\prob\left(C_{\rm sym,Ala}(P)<R\right)}{\prob\left(C_{\rm sym,opt}(P)<R\right)}\leq16,
\end{align}
and
\begin{align}
    \liminf_{P \rightarrow \infty} \frac{\prob\left(C_{\rm sym,Ala}(P)<R\right)}{\prob\left(C_{\rm sym,opt}(P)<R\right)}\geq16.
\label{eq:latter}
\end{align}

We start by showing the former inequality.
By \eqref{eq:csim_opt}, for any  rate $R$,
\begin{align}
    & \prob\left(C_{\rm sym,opt}(P)<R\right) \nonumber \\
    &  = \prob\left(\min \left\{ C_{\rm opt}(\{1\},P),C_{\rm opt}(\{2\},P),C_{\rm opt}(\{1,2\},P)\right\} < R\right)  \nonumber  \\
    & \overset{\text{\rm (a)}}\geq \prob\left(\bigcup_{k=1,2} \left\{ C_{\rm opt}(\{k\},P)< R \right\} \right) \nonumber \\
    & \overset{\text{\rm (b)}}= \sum_{k=1}^2\prob\left(C_{\rm opt}(\{k\},P)< R\right),
    \label{eq:boundOfOpt}
\end{align}
where $(a)$ follows since we are taking  into account only the events in which a single-user constraint constitutes the bottleneck  and $(b)$ follows since the fading coefficients are independent and hence so are the corresponding single-user constraint events.

As for the outage probability of diversity combining, invoking \eqref{eq:csim_ala}, we may upper bound it by applying the union bound:
\begin{align}
    & \prob\left(C_{\rm sym,Ala}(P)<R\right) \nonumber \\
    &  = \prob\left(\min \left\{ C_{\rm Ala}(\{1\},P),C_{\rm Ala}(\{2\},P),C_{\rm Ala}(\{1,2\},P)\right\} < R\right)  \nonumber \\
    & \leq \sum_{k=1}^2\prob\left(C_{\rm Ala}(\{k\},P)< R\right) + \prob\left(C_{\rm Ala}(\{1,2\},P)<R\right).
    \label{eq:boundOfAla}
\end{align}

For  a general $N_r\times N_t$ i.i.d. Rayleigh fading channel between a user and receiver, the Frobenius norm squared $\|\svv{H}_k\|^2_{F}$ is chi-square distributed with  $2 N_t N_r$ degrees of freedom. Further, it is readily shown that for small $\epsilon$
\begin{align}
    \prob\left(\|\svv{H}_k\|^2_{F}<\epsilon\right)= c(N_t N_r) \cdot \epsilon ^{N_t N_r} + o\left(\epsilon^{N_t N_r}\right),
    \label{eq:probOfChi}
\end{align}
where $o(\epsilon)/\epsilon \rightarrow 0$ as $\epsilon \rightarrow 0$ and where $c(N_t N_r)$ is a constant that depends only on $N_t N_r$.\footnote{It can be shown, e.g., that for $N_t N_r=4$, $c(4)=\frac{1}{384}$.}

In the case considered of $1 \times 2$ channel matrices, $\|\svv{H}_k\|^2_{F}=\|\svv{h}_k\|^2$ and  $N_t N_r=2$. Therefore,
\begin{align}
\prob\left(C_{\rm opt}(\{k\},P)<R\right) &= \prob\left(\log\left(1+\PowerMat\|{\bf h}_k\|^2\right)<R\right) \nonumber \\
& = \prob\left(\|{\bf h}_k\|^2<\frac{2^R-1}{P}\right) \nonumber \\
& = c(4) \left(\frac{2^R-1}{P}\right)^4 +o\left(\frac{1}{P^4}\right).
\label{eq:aymptot_last}
\end{align}
Similarly,
\begin{align}
    \prob\left(C_{\rm Ala}(\{k\},P)<R\right)& = c(4)\left(\frac{2^R-1}{P/2}\right)^4+o\left(\frac{1}{P^4}\right)
\end{align}
    and
\begin{align}
    \prob\left(C_{\rm Ala}(\{1,2\},P)<R\right)&
    =c(8)\left(\frac{2^R-1}{P/2}\right)^8 +o\left(\frac{1}{P^8}\right).
\end{align}

Combining these asymptotics with (\ref{eq:boundOfOpt}) and (\ref{eq:boundOfAla}) yields
\begin{align}
    &\frac{\prob\left(C_{\rm sym,Ala}(P)<R\right)}{\prob\left(C_{\rm sym,opt}(P)<R\right)} \nonumber \\
    &\leq \frac{\displaystyle{\sum_{k=1}^2\prob\left(C_{\rm Ala}(\{k\},P)< R\right) + \prob\left(C_{\rm Ala}(\{1,2\},P)<R\right)}}    {\displaystyle{\sum_{k=1}^2\prob\left(C_{\rm opt}(\{k\},P)< R\right)}} \nonumber \\
    & = \frac{2c(4)\left(\frac{2^R-1}{P/2}\right)^4
    +o\left(\frac{1}{P^4}\right)
    +c(8)\left(\frac{2^R-1}{P/2}\right)^8
    +o\left(\frac{1}{P^8}\right)
    }
    {2c(4)\left(\frac{2^R-1}{P}\right)^4+o\left(\frac{1}{P^4}\right)}
    \nonumber \\
    & \xrightarrow{P \rightarrow \infty} 16.
\end{align}
Hence
\begin{align}
    \limsup_{P \rightarrow \infty}\frac{\prob\left(C_{\rm sym,Ala}(P)<R\right)}{\prob\left(C_{\rm sym,opt}(P)<R\right)} \leq 16.
\label{eq:limsup1}
\end{align}

We turn now to establishing \eqref{eq:latter}.
Applying the same arguments as in \eqref{eq:boundOfOpt} to $C_{\rm sym,Ala}(P)$, we get
\begin{align}
     \prob\left(C_{\rm sym,Ala}(P)<R\right)
    & \geq \sum_{k=1}^2\prob\left(C_{\rm Ala}(\{k\},P)< R\right).
    \label{eq:boundOfOpt_appliedOnAla}
\end{align}
Applying \eqref{eq:boundOfAla} to $C_{\rm sym,opt}(P)$, we get
\begin{align}
    & \prob\left(C_{\rm sym,opt}(P)<R\right) \nonumber \\
    & \leq \sum_{k=1}^2\prob\left(C_{\rm opt}(\{k\},P)< R\right) + \prob\left(C_{\rm opt}(\{1,2\},P)<R\right).
\end{align}
Recalling \eqref{eq:csim_opt} and Eq. (5) from \cite{sandhu2000space}, we have
\begin{align}
    C_{\rm opt}(\{1,2\}) &= \log\det\left(\svv{I}+\PowerMat \cdot \svv{H}_{\rm comb}\svv{H}_{\rm comb}^H \right) \nonumber \\
    &\geq \log\left(1+\PowerMat \|\svv{H}_{\rm comb}\|_F^2\right)
\end{align}
and hence
\begin{align}
    \prob\left(C_{\rm opt}(\{1,2\},P)<R\right) & \leq \prob(\log\left(1+\PowerMat \|\svv{H}_{\rm comb}\|_F^2\right)<R) \nonumber \\
    & = \prob\left(\|\svv{H}_{\rm comb}\|_F^2 < \frac{2^R-1}{P} \right) \nonumber \\
    & = c(8)\left(\frac{2^R-1}{P}\right)^8 +o\left(\frac{1}{P^8}\right).
\end{align}
It follows that
\begin{align}
     & \prob\left(C_{\rm sym,opt}(P)<R\right)  \nonumber \\ 
     & \leq 2c(4)\left(\frac{2^R-1}{P}\right)^4
    +o\left(\frac{1}{P^4}\right)
    +c(8)\left(\frac{2^R-1}{P}\right)^8
    +o\left(\frac{1}{P^8}\right)
    \label{eq:boundOfAla_appliedOnOpt}
\end{align}

Combining \eqref{eq:boundOfOpt_appliedOnAla} and \eqref{eq:boundOfAla_appliedOnOpt}, we obtain
\begin{align}
    &\frac{\prob\left(C_{\rm sym,Ala}(P)<R\right)}{\prob\left(C_{\rm sym,opt}(P)<R\right)} \nonumber \\
    &\geq \frac{\displaystyle{\sum_{k=1}^2\prob\left(C_{\rm Ala}(\{k\},P)< R\right)}}
    {\displaystyle{\sum_{k=1}^2\prob\left(C_{\rm opt}(\{k\},P)< R\right) + \prob\left(C_{\rm opt}(\{1,2\},P)<R\right)}}     \nonumber \\
    & = \frac{2c(4)\left(\frac{2^R-1}{P}\right)^4+o\left(\frac{1}{P^4}\right)}
    {2c(4)\left(\frac{2^R-1}{P}\right)^4
    +o\left(\frac{1}{P^4}\right)
    +c(8)\left(\frac{2^R-1}{P}\right)^8
    +o\left(\frac{1}{P^8}\right)} \nonumber \\
    & \xrightarrow{P \rightarrow \infty} 16.
\end{align}
Hence,
\begin{align}
    \liminf_{P \rightarrow \infty}\frac{\prob\left(C_{\rm sym,Ala}(P)<R\right)}{\prob\left(C_{\rm sym,opt}(P)<R\right)} \geq 16.
    \label{eq:liminf1}
\end{align}
Combining \eqref{eq:limsup1} and \eqref{eq:liminf1}, we have established \eqref{eq:limAla}.

Next, we show that a reduction by a factor of two  in transmission power asymptotically as $P\rightarrow \infty$ translates to an increase by the same factor of $16$ in outage probability, i.e.
\begin{align}
    \lim_{P \rightarrow \infty} \frac{\prob\left(C_{\rm sym,opt}(P/2)<R\right)}{\prob\left(C_{\rm sym,opt}(P)<R\right)}=16.
    \label{eq:powerfactor}
\end{align}

To this end, by the union bound, we have
\begin{align}
    & \prob\left(C_{\rm sym,opt}(P/2)<R\right) \nonumber \\
    & \leq \sum_{k=1}^2\prob\left(C_{\rm opt}(\{k\},P/2)< R\right) + \prob\left(C_{\rm opt}(\{1,2\},P/2)<R\right),
\end{align}
which combined with \eqref{eq:boundOfOpt}, implies that
\begin{align}
    &\frac{\prob\left(C_{\rm sym,opt}(P/2) < R\right)}{\prob\left(C_{\rm sym,opt}(P) < R\right)} \nonumber \\
    & \leq \frac{2c(4)\left(\frac{2^R-1}{P/2}\right)^4+o\left(\frac{1}{P^4}\right)+c(8)\left(\frac{2^R-1}{P/2}\right)^8+o\left(\frac{1}{P^8}\right)}{2c(4)\left(\frac{2^R-1}{P}\right)^4+o\left(\frac{1}{P^4}\right)}
    \nonumber \\
    &\xrightarrow{P \rightarrow \infty} 16,
\end{align}
where we have used the asymptotics \eqref{eq:boundOfAla_appliedOnOpt} for the denominator and 
Hence,
\begin{align}
    \limsup_{P \rightarrow \infty}\frac{\prob\left(C_{\rm sym,opt}(P/2) < R\right)}{\prob\left(C_{\rm sym,opt}(P) < R\right)} \leq 16.
    \label{eq:limsup2}
\end{align}

Next, from \eqref{eq:boundOfOpt}, we have that
\begin{align}
    & \prob\left(C_{\rm sym,opt}(P/2)<R\right) \geq \sum_{k=1}^2\prob\left(C_{\rm opt}(\{k\},P/2)< R\right),
\end{align}
which combined  with \eqref{eq:boundOfAla_appliedOnOpt}, implies that 
%
\begin{align}
    &\frac{\prob\left(C_{\rm sym,opt}(P/2) < R\right)}{\prob\left(C_{\rm sym,opt}(P) < R\right)} \nonumber \\
    & \geq \frac{2c(4)\left(\frac{2^R-1}{P/2}\right)^4+o\left(\frac{1}{P^4}\right)}{2c(4)\left(\frac{2^R-1}{P}\right)^4+o\left(\frac{1}{P^4}\right)+c(8)\left(\frac{2^R-1}{P}\right)^8+o\left(\frac{1}{P^8}\right)}
    \nonumber \\
    & \xrightarrow{P \rightarrow \infty} 16.
\end{align}
Hence, 
\begin{align}
    \liminf_{P \rightarrow \infty}\frac{\prob\left(C_{\rm sym,opt}(P/2) < R\right)}{\prob\left(C_{\rm sym,opt}(P) < R\right)} \geq 16.
    \label{eq:liminf2}
\end{align}
Combining \eqref{eq:limsup2} and \eqref{eq:liminf2}, we have established 
\eqref{eq:powerfactor}.

Therefore, optimal detection for transmission at half the power yields the same outage probability as that achieved with diversity combining, at asymptotically high SNR.


We conclude by noting the proof of the claim for a general number of users follows along similar lines. This can be seen by noticing that the dominant terms in the outage probability (for asymptotically high SNR) are the ones corresponding to single-user constraints. The latter can be approximated at high SNR using (\ref{eq:probOfChi}), tracing the same steps as for the two-user case.
\bibliographystyle{IEEEtran}
\bibliography{eladd}

\begin{thebibliography}{10}
\providecommand{\url}[1]{#1}
\csname url@samestyle\endcsname
\providecommand{\newblock}{\relax}
\providecommand{\bibinfo}[2]{#2}
\providecommand{\BIBentrySTDinterwordspacing}{\spaceskip=0pt\relax}
\providecommand{\BIBentryALTinterwordstretchfactor}{4}
\providecommand{\BIBentryALTinterwordspacing}{\spaceskip=\fontdimen2\font plus
\BIBentryALTinterwordstretchfactor\fontdimen3\font minus
  \fontdimen4\font\relax}
\providecommand{\BIBforeignlanguage}[2]{{%
\expandafter\ifx\csname l@#1\endcsname\relax
\typeout{** WARNING: IEEEtran.bst: No hyphenation pattern has been}%
\typeout{** loaded for the language `#1'. Using the pattern for}%
\typeout{** the default language instead.}%
\else
\language=\csname l@#1\endcsname
\fi
#2}}
\providecommand{\BIBdecl}{\relax}
\BIBdecl

\bibitem{venkataramani2000perfect}
R.~Venkataramani and Y.~Bresler, ``Perfect reconstruction formulas and bounds
  on aliasing error in sub-{N}yquist nonuniform sampling of multiband
  signals,'' \emph{IEEE Transactions on Information Theory}, vol.~46, no.~6,
  pp. 2173--2183, 2000.

\bibitem{brennan1959linear}
D.~G. Brennan, ``Linear diversity combining techniques,'' \emph{Proceedings of
  the IRE}, vol.~47, no.~6, pp. 1075--1102, 1959.

\bibitem{molisch2004mimo}
A.~F. Molisch and M.~Z. Win, ``{MIMO} systems with antenna selection,''
  \emph{IEEE Microwave Magazine}, vol.~5, no.~1, pp. 46--56, 2004.

\bibitem{sanayei2004antenna}
S.~Sanayei and A.~Nosratinia, ``Antenna selection in {MIMO} systems,''
  \emph{IEEE Communications Magazine}, vol.~42, no.~10, pp. 68--73, 2004.

\bibitem{alamouti1998simple}
S.~M. Alamouti, ``A simple transmit diversity technique for wireless
  communications,'' \emph{IEEE Journal on Selected Areas in Communications},
  vol.~16, no.~8, pp. 1451--1458, 1998.

\bibitem{vishwanath2003duality}
S.~Vishwanath, N.~Jindal, and A.~Goldsmith, ``Duality, achievable rates, and
  sum-rate capacity of gaussian {MIMO} broadcast channels,'' \emph{IEEE
  Transactions on Information Theory}, vol.~49, no.~10, pp. 2658--2668, 2003.

\bibitem{el2007relay}
A.~El~Gamal, N.~Hassanpour, and J.~Mammen, ``Relay networks with delays,''
  \emph{IEEE Transactions on Information Theory}, vol.~53, no.~10, pp.
  3413--3431, 2007.

\bibitem{khormuji2010instantaneous}
M.~N. Khormuji and M.~Skoglund, ``On instantaneous relaying,'' \emph{IEEE
  Transactions on Information Theory}, vol.~56, no.~7, pp. 3378--3394, 2010.

\bibitem{heath2001antenna}
R.~W. Heath, S.~Sandhu, and A.~Paulraj, ``Antenna selection for spatial
  multiplexing systems with linear receivers,'' \emph{IEEE Communications
  letters}, vol.~5, no.~4, pp. 142--144, 2001.

\bibitem{liu2015optimized}
L.~Liu and R.~Zhang, ``Optimized uplink transmission in multi-antenna {C-RAN}
  with spatial compression and forward,'' \emph{IEEE Transactions on Signal
  Processing}, vol.~63, no.~19, pp. 5083--5095, 2015.

\bibitem{tirkkonen2000minimal}
O.~Tirkkonen, A.~Boariu, and A.~Hottinen, ``Minimal non-orthogonality rate 1
  space-time block code for 3+ {Tx} antennas,'' in \emph{Spread Spectrum
  Techniques and Applications, 2000 IEEE Sixth International Symposium on},
  vol.~2.\hskip 1em plus 0.5em minus 0.4em\relax IEEE, 2000, pp. 429--432.

\bibitem{jafarkhani2001quasi}
H.~Jafarkhani, ``A quasi-orthogonal space-time block code,'' \emph{IEEE
  Transactions on Communications}, vol.~49, no.~1, pp. 1--4, 2001.

\bibitem{sharma2003improved}
N.~Sharma and C.~B. Papadias, ``Improved quasi-orthogonal codes through
  constellation rotation,'' \emph{IEEE Transactions on Communications},
  vol.~51, no.~3, pp. 332--335, 2003.

\bibitem{tarokh1999space}
V.~Tarokh, H.~Jafarkhani, and A.~R. Calderbank, ``Space-time block codes from
  orthogonal designs,'' \emph{IEEE Transactions on Information Theory},
  vol.~45, no.~5, pp. 1456--1467, 1999.

\bibitem{adams2011novel}
S.~S. Adams, J.~Davis, N.~Karst, M.~K. Murugan, B.~Lee, M.~Crawford, and
  C.~Greeley, ``Novel classes of minimal delay and low {PAPR} rate 1/2 complex
  orthogonal designs,'' \emph{IEEE Transactions on Information Theory},
  vol.~57, no.~4, pp. 2254--2262, 2011.

\bibitem{domanovitz2017diversity}
\BIBentryALTinterwordspacing
E.~Domanovitz and U.~Erez, ``Diversity combining via universal orthogonal
  space-time transformations.'' [Online]. Available:
  \url{http://www.eng.tau.ac.il/\%7Euri/universal_combining.pdf}
\BIBentrySTDinterwordspacing

\bibitem{sandhu2000space}
S.~Sandhu and A.~Paulraj, ``Space-time block codes: A capacity perspective,''
  \emph{IEEE Communications Letters}, vol.~4, no.~12, pp. 384--386, 2000.

\end{thebibliography}

\end{document}